\documentclass[onecolumn,12pt]{asme2ej}
\pdfoutput=1

\usepackage{setspace}
\usepackage{epsfig} 
\usepackage{multirow}
\usepackage{amsmath}
\usepackage{pdflscape}
\usepackage[outdir=./]{epstopdf}

\usepackage{lastpage}
\usepackage{fancyhdr}
\usepackage{lastpage}
\usepackage{xfrac}
\usepackage{booktabs}
\usepackage{caption} 
\usepackage[modulo]{lineno}
\usepackage{subfig}
\usepackage{xcolor}
\usepackage[para]{threeparttable}

\usepackage{booktabs}
\usepackage{caption} 
\newcommand{\ds}{\displaystyle}

\newif\ifproofread

\newcommand{\new}[1]{%
\ifproofread
\textcolor{red}{#1}%
\else
#1%
\fi
}

\title{Prediction of separation and transition on a low-pressure turbine blade using a RANS grid}

\author{Rajesh Ranjan\thanks{Address all correspondence to this author.}
    \affiliation{
	Dept. of Mechanical \& Aerospace Engg.\\
	The Ohio State University, USA\\
    Email: ranjan.25@osu.edu
    }	
}

\author{S. M. Deshpande
            \affiliation{ Engineering Mechanics Unit\\
            	JNCASR, Bangalore, India\\
    Email: smd@jncasr.ac.in
    }
    }

\author{Roddam Narasimha
            \affiliation{ Engineering Mechanics Unit\\
            	JNCASR, Bangalore, India\\
    Email: roddam@jncasr.ac.in
    }
}

\begin{document}
\proofreadfalse  

\maketitle    
\today
\newpage
\begin{abstract}
{\it 
Flow past a high-lift low-pressure turbine (LPT) blade in a cascade could be quite complex  as phenomena like separation and transition are often involved.
For a highly loaded T106A blade at high incidence and relatively low Reynolds number ($25,000 < Re < 1,00,000$), separation-induced transition is observed on the suction side of the blade, making it a challenging problem for model-based simulations.
In this work, computations for this flow are carried out using RANS and hybrid LES/RANS approaches. The RANS simulations are performed with six popular low-$Re$ turbulence models. 
\new{
While turbulence models by themselves fail to predict any separation on T106A blade,  the four-equation Langtry-Menter transition model predicts a short separation bubble.
The characteristic of this bubble, however, is very different from what is observed in experiments and DNS, and therefore transition is not accurately predicted.  
An embedded hybrid LES/RANS approach, Limited numerical scales (LNS), with an automatic switch to LES in sufficiently resolved grids, is then used for predictions on the same RANS grid.
With the statistical turbulence on fine grids, LES-like behavior of LNS results in an unphysical drop in Reynolds stresses as the turbulent fluctuations are not appropriately represented on the resolved scale. 
Therefore, the LNS results are very similar to those obtained with  turbulence models. 
However, when synthetic turbulence with correct statistical characteristics is used to stimulate the large eddies in the embedded LES zone, LNS is able to predict separation and recovers a solution very close to DNS and experimental results.
This shows that the inherent unsteadiness in the LPT flows at transitional conditions is not sufficient to sustain the turbulence in the LES region, and therefore a model is necessary to  enable a transfer between statistical and directly-resolved kinetic energies during RANS/LES switch.
Finally, the present work demonstrates that affordable solutions for transitional flows on turbine blades can be found on RANS grids if proper models are used, rather than taking a complete recourse to expensive methods such as LES or DNS.
}
}
\end{abstract}
\newpage
\begin{nomenclature}
\entry{$\ds U$}{~~Velocity}
\entry{$\ds \rho$}{~~~Density}
\entry{$\ds \nu$}{~~~Kinematic viscosity}
\entry{$\ds Re$}{~Reynolds number}
\entry{$\ds \theta$}{~~~Momentum thickness}
\entry{$\ds M_{is}$}{Isentropic Mach number}
\entry{$\ds L_{ax}$}{Axial chord length of the blade}
\entry{$\ds \beta_{1}$}{~Angle of incidence}
\entry{$k$}{~~~Turbulence kinetic energy}
\entry{$\bar{k}$}{~~~Subgrid-scale turbulence kinetic energy}
\entry{$\epsilon$} {~~~Turbulent dissipation rate}
\entry{$\omega$}{~~Turbulent frequency}
\entry{$\bar{v}_2$}{~~Energy of the fluctuating velocity components normal to the streamlines}
\entry{$\nu_t$} {~~Eddy viscosity}
\entry{$\Tilde{\nu}$} {~~~Spalart–Allmaras variable}
\entry{$R_t$} {~Turbulent Reynolds number}
\entry{$\gamma$} {~~~Intermittency}
\entry{$\ds \beta_{1}$}{~Blade angle of incidence}
\entry{$\ds p_{t1}$}{Total inlet pressure}
\entry{$\ds p_{ref}$}{Reference pressure}
\entry{$\ds c_f$}{~~Coefficient of friction}
\entry{$\ds c_p$}{~~Coefficient of pressure}
\entry{$\Delta y^+$}{Grid spacing in wall-normal direction in wall units}
\end{nomenclature}

\newpage
\section{Introduction}
Low-pressure turbine (LPT) blades are generally highly loaded to minimize the weight of the turbine by reducing the number of blades per stage as well as the number of such turbine stages. 
On the other hand, higher aerodynamic loading can result in higher
overall losses, which in turn can lead to a decrease in the stage efficiency.  
Particularly in cruise conditions at high altitudes, these issues become important when the flow in the LP turbines is at low Reynolds numbers.
The loading on the suction side in these situations may lead to a
 favorable  pressure gradient followed by a strong adverse pressure gradient, causing the stable boundary layer to separate.
Based on the extent of separation, which may or may not attach, the loss due to pressure can be significant. 
Thus, it is imperative to obtain a clear description of flow around LPT blades during such situations to achieve a reliable design. 

There have been several significant experimental efforts to understand these flows\cite{stadtmuller1,hodson2007role, schobeiri2014comparative} on different LPT blades.
All these measurements indicate laminar boundary layer separation on the suction side under certain situations as discussed above.
However, obtaining a completely resolved three-dimensional flow field, along with an accurate description of onset of separation as well as extent of separation bubble and associated losses through measurements, are often challenging because of limitations of the instruments used in the experiments.

Computational Fluid Dynamics (CFD) on the other hand can provide a detailed flow field along with wall fluxes, provided all the pertinent scales are properly resolved. 
Indeed, direct numerical simulations (DNS) in the past\cite{kalitzin2003dns, sandberg2015compressible, ranjan2017new}  have given significant insights into these flows, and have enriched our understanding. 
However these simulations are computationally very expensive, and  are   used rarely for design. 
A relatively less expensive approach is Large Eddy Simulation (LES), where large eddies are simulated whereas smaller eddies are modelled using subgrid-scale models. 
This approach has also been used in several studies such as those of  Michelassi \textit{et al.}\cite{michelassi2002}, but the computational costs of these simulations are also rather high and cannot be used routinely for parameter studies. 

To predict these flows, gas turbine industries often rely on Reynolds-averaged Navier-Stokes (RANS) solvers with transition models. 
RANS has been very successful in predicting a large class of engineering flows, but their success rate in predicting transitional flow with separation and reattachment is very underwhelming \cite{wauters2018study}. 
In fact, a review of  transition models for turbomachinery flow\cite{dick2017transition} found no conclusive evidence that shows these models as reliable.
The difficulty in predicting  transition using RANS is inherent in the averaged equations, which neglect the amplification of disturbances of specific frequencies that cause transition. 
Thus, the models rely on combining different methods to predict separation as accurately as possible over a range of different transition mechanisms and conditions.
Wauters \& Degroote\cite{wauters2018study}, and Dick \& Kubacki \cite{dick2017transition} present excellent reviews of modern turbulence and transition models for interested readers. 

Because of the reasons mentioned above, more and more industries are now adopting strategies of hybrid RANS/LES models for complex applications. 
These approaches rely on LES for resolving critical regions such as that near separation, while the rest of the domains is solved using RANS. 
Computationally these methods are very affordable, as the grid requirements are often very similar to those of RANS\cite{batten2002lns}. Coupling of RANS and LES approaches is generally performed under two main categories: zonal models and embedded models. 
While in the former the zones of applicability of LES and RANS are well defined, no boundary exists between the RANS and LES parts of the flow in the latter approach. 

Two reviews \cite{frohlich2008hybrid, labourasse2004advance} present excellent summaries of different hybrid models. 
Briefly, the most popular approaches are Detached Eddy Simulation (DES; \cite{spalart1997comments}),  Limited-Numerical-Scales (LNS; \cite{batten2002lns}) and their variants. In DES, the transport equation for the eddy viscosity is manipulated in such a way that near the wall a RANS model is used, while a subgrid-scale (SGS) model takes over away from the wall. In LNS, the Reynolds stress term is treated in such as way that the RANS model recovers LES-like behavior when the grid spacing becomes small. 
While DES and their variants are the most widely used hybrid models, LNS has shown promising results in many complex transitional flows around industrial geometries on moderate grids\cite{labourasse2004advance}. Examples include flow over a periodic hill\cite{batten2002lns}, separation control of the flow over a blunt aerofoil\cite{Chakravarthy2002}, and isothermal flow in a ventilated room\cite{jouvray2005computation}.



In the present work, RANS and hybrid RANS/LES simulations are performed for flow around the highly loaded high-lift LPT  blade T106A in a cascade.
The simulation conditions are based on the experiments carried out \cite{stadtmuller1} on the widely used Pratt and Whitney blade T106A at a sub-critical Reynolds number $Re = 51,831$ (based on the inlet velocity and the axial-chord) and a relatively high angle of incidence ($\beta_1 = 45.5^0$).
Under these conditions, the flow results in a separation-induced transition towards the trailing edge of the suction side as described in the high-resolution DNS studies\cite{ranjan2017new, ranjan2015advances, sandberg2015compressible}.

Table~\ref{les_lpt} lists some of the major RANS studies on this blade at similar sub-critical flow conditions. 
\begin{table*}
\caption{RANS studies on T106A blade}
\begin{center}
\label{les_lpt}
\centering 

\def\arraystretch{1.6}
\begin{tabular}{|p{3cm}| l| l| l| l| l|l|}

\toprule

\multirow{2}{*}{\bfseries{Study}} & \bfseries{Turbulence} & \bfseries{${Re}$} & \multicolumn{4}{|c|}{\bfseries{Grid}}\\

\cline{3-7}
&\bfseries{Models}&$(\times 10^4)$& $(N_x\times N_y)$ &$N_z$&Size ($\times10^6$)&$\Delta Y^+$\\

\midrule

Stadtmuller \cite{stadtmuller1} & -  & 5.18 & -  &- &-&- \\
\hline

Michelassi \textit{et. al}\cite{michelassi2002} & $k$-$\omega$  &5.18 & $256\times144$  & NA & 0.04  & -\\
\hline

Marciniak \textit{et. al} \cite{marciniak2010predicting} & $k$-$\omega$, $\gamma-Re_{\theta}$ (2D)  & $15-110$ & 19,290 &  NA & 0.02 &  1 \\
\hline

Akolekar \textit{et. al} \cite{akolekar2019development} & $k$-$\omega$ SST, SA, $\gamma$-$Re_{\theta}$, $k\bar{v}_2\omega$, LKE & 5.18 & 2,70,000 & 336 & 18.1 & 0.4 \\
\hline

Present & See Table~\ref{tab:models}   & 5.18 & 77,303 & 32 &   2.5 &1\\

\bottomrule
\end{tabular}
\end{center}
\end{table*}
The first RANS study was performed by Stadtmuller \cite{stadtmuller1}, who also conducted experiments with the objective of estimating the `real' flow angle, as the geometric angle measured in the experiment was uncertain due to the presence of upstream wake-generating bars. They also studied the effect of free-stream turbulence levels on the flow.
Michelassi  \textit{et al.} \cite{michelassi2002} carried out unsteady RANS (URANS) simulations using the $k$-$\omega$ model, with a transition model proposed by Mayle\cite{Mayle1991}, which incorporated with the effect of the periodic incoming wake . 
They\cite{michelassi2002} also found an early transition to turbulence compared to their DNS, and hence models needed to be tuned based on information from DNS to improve the URANS prediction.

Marciniak \textit{et al.} \cite{marciniak2010predicting} have used the updated DLR-TRACE RANS code to study the effect of turbulence models on predicting separation-induced transition on the T106A blade over a $Re$ range of $1.5\times10^5-11\times10^5$, based on the exit velocity and chord length, and at a slightly lower angle of incidence $\beta_1=37.7^{\circ}$. At low values of $Re(< 5\times 10^5)$, where the boundary layer is laminar on most of the blade, the fully turbulent simulation with the Wilcox k-$\omega$ model fails to predict any separation. However their simulations with the Langtry-Menter $\gamma\!-\!Re_{\theta}$ transition model as well as their in-house multimode transition model show improved prediction of separation, though losses are overestimated. At high Reynolds numbers, the boundary layer is turbulent and no separation is observed: these flows are well-predicted by all the turbulence models.

Recently, Akolekar \textit{et al.}\cite{akolekar2019development}  performed RANS simulations with five turbulence models using OPENFOAM but on a highly resolved grid previously used for DNS \cite{sandberg2015compressible}.
No model except the laminar kinetic energy (LKE)\cite{lopez2016prediction} and $k\bar{v}_2\omega$\cite{walters2008three} models predicts separation near the trailing edge of the suction side. The $\gamma\!-\!Re_{\theta}$ model also fails to predict separation in their simulation. 

Table \ref{les_lpt} is not exhaustive and is given only for reference purposes.
Recent RANS studies, performed for transitional flows on other LPT blades, such as T106C\cite{minot2016improvement, marciniak2015modeling}, T106D\cite{lodefier2006modelling}, MTU\cite{muller2016dns} and  Pak B\cite{schobeiri2014comparative, nikparto2016numerical}, also describe the difficulties that RANS models face in predicting flow behavior.
The flow on highly loaded turbine blades at low Reynolds numbers, when the boundary layer is attached on most of the blade surface and separation is observed only near the trailing edge on the suction side, is a particularly challenging case for RANS.
On the other hand, for hybrid RANS/LES models, the challenge may be due to not having sufficient inherent unsteadiness in the flow (such as in massively separated base or wake flows) in order to drive turbulence in the LES regions. 
Thus the large-scale, statistical turbulence kinetic energy (TKE)
as  obtained by RANS equations may quickly dissipate in this region unless eddies are stimulated artificially. 

The present paper addresses these crucial issues by providing a critical comparison of various RANS and hybrid RANS/LES models. 
All RANS turbulence models employed are either low-$Re$ or have an eddy viscosity that is specially treated to include near-wall effects. 
The hybrid RANS/LES simulation is performed  with LNS approach in order to resolve the unsteady scales near the separation region.
The issue of consistent treatment of TKE in RANS/LES regions, in order to avoid unphysical behavior, is also addressed through artificial stimulation of large eddies in the LES domain.
The implementation of the LNS model for transitional flows over LPTs is here performed for the first time, to the best of our knowledge.

\section{Numerical Setup}
\subsection{Computational Domain and Grid}
In the present study, the simulations are carried out  using the commercial  code CFD++ version 15.1 by Metacomp Technologies \cite{chakravarthy1999cfd++}. The flow is simulated using preconditioned compressible Navier-Stokes equations in a finite volume framework.  The equations are solved in dimensional form as shown in Table~\ref{tab:flow_parameters}. These values are taken from the experiment\cite{stadtmuller1} for the steady case (i.e. no upstream wake), and roughly give the inlet and exit Reynolds numbers based on the axial chord as 51,831 and 60,535 respectively. At the inlet, total pressure and temperature are specified along with the flow direction. At the outlet, back pressure is imposed using characteristics-based boundary conditions. Isothermal no-slip boundary conditions are applied on the blade surface. All the simulations were performed with a free-stream turbulence level of $2.2\%$, as used by Stadtmuller\cite{stadtmuller1} for the RANS computations with the DLR-TRACE code.  

\begin{table}[]
    \caption{Flow parameters for current simulations}
    \centering
\begin{tabular}{|c|c|}
\toprule
\textbf{Parameter} & \textbf{Value} \\
\midrule
    Inlet flow angle $\beta_1$  & $45.5^{\circ}$  \\
    \hline
    Total pressure at inlet $p_{t1}$ (Pa) & 7,770  \\
        \hline
    Static pressure at inlet $p_{1}$ (Pa) & 7,340 \\
        \hline
    Static pressure at exit $p_2$ (Pa) & 6,950 \\
        \hline
    Total temperature at inlet    $T_{t1}$ (K) & 312.9  \\
    \bottomrule
\end{tabular}
    \label{tab:flow_parameters}
\end{table}

\begin{figure}
\centering
\includegraphics[trim=1cm 7cm 3cm 1cm, clip, width=0.95\linewidth, angle=0]{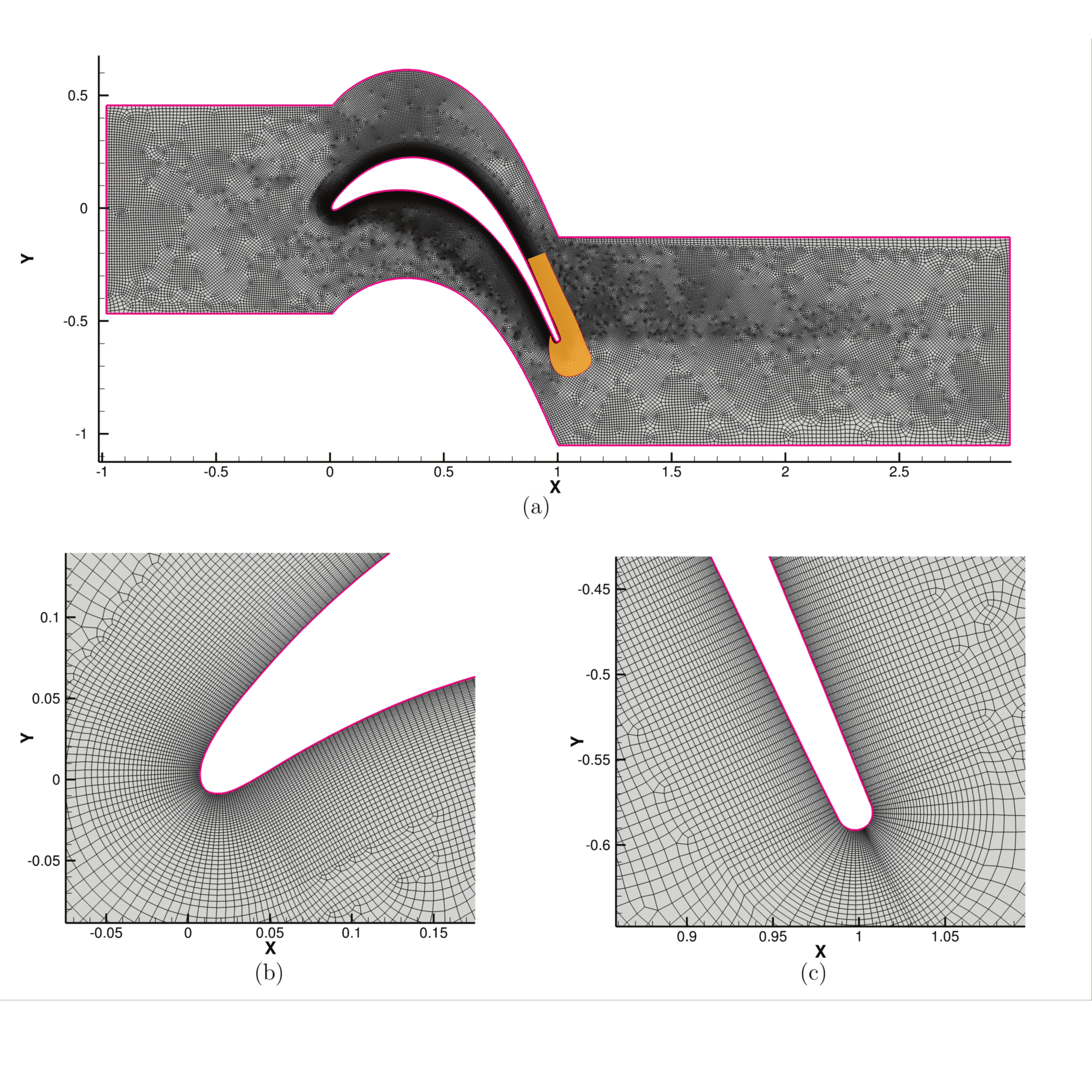}
  \caption{Computational domain and grid used for RANS and LNS. Darker regions in (a) represent high grid density. Brown shaded-region show the unsteady separation region at current flow conditions where high grid density ensures that the stresses are computed using SGS approach.}
  \label{fig:domain_lns}
\end{figure}

All RANS and LNS simulations are performed in a three-dimensional setup. Though the flow is homogeneous in the spanwise direction and could be performed in 2D for pure RANS simulations, a 3D domain is necessary in LNS to account for spanwise fluctuations in LES zones. In order to assess the capability of LNS to predict critical flow features with reasonable resources, the computations are performed on the RANS grid as shown in Table \ref{les_lpt}. This is similar to the approach of Batten \cite{batten2002lns}, who used the same grids for both RANS and LNS for flow over a periodic hill, and obtained much more accurate solutions for the latter case. The computational domain and the grid used for  simulations are  shown in Fig. \ref{fig:domain_lns}. 
The dimension in the periodic pitch-wise direction is one blade-spacing, while inflow and outflow boundaries are located $L_{ax}$ upstream of the leading edge and $2L_{ax}$ downstream of the trailing edge respectively. This is to avoid reflection of any spurious waves at these boundaries as no buffer layer is used in these simulations. The spanwise width is kept at $0.2L_{ax}$, chosen based on prior LES studies\cite{michelassi2002} at the same or greater $Re$.


The computational domain consists of hexahedral elements with a boundary layer grid near the blade as shown in Fig.~\ref{fig:domain_lns}(a). 
Enlarged views of grids near the leading and trailing edges are shown in Fig.~\ref{fig:domain_lns}(b) and \ref{fig:domain_lns}~(c) respectively. 
There are 740 points around the blade to capture the leading and trailing edge curvatures sufficiently well.
The grid topology followed is similar to that used in the DNS study by the authors\cite{ranjan2017new}.
Thus, the total grid size used for the present simulations is 2.5 million elements, of which about half are present in the boundary layer.
Grid independence of the solution is ensured for both RANS and LNS cases by comparing the results with double the chosen grid size. 

\begin{figure}
\centering
\includegraphics[trim=0 0 0 0, clip, width=0.95\linewidth]{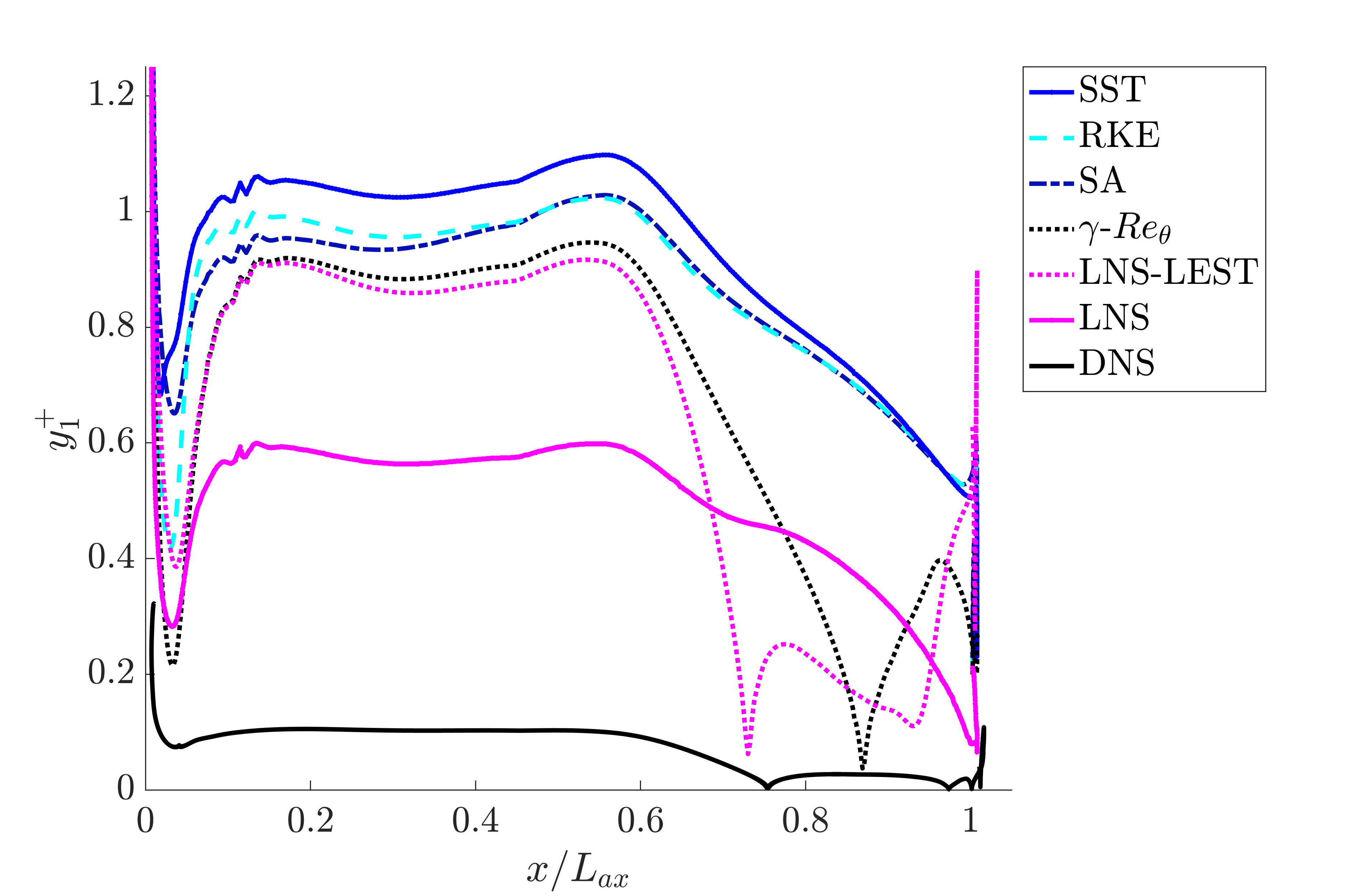}
\caption{$y_1^{+}$ on the suction side of the blade. DNS data is from Ranjan \textit{et al.}\cite{ranjan2017new} on a 161 million grid.}
\label{fig:yp_rans}
\end{figure}

Figure \ref{fig:yp_rans} shows the distribution of the first-cell distance in wall-units ($y_1^{+}$) on the suction side of the blade.
This value is maintained below 1 for most part of the blade for the models, and hence a solve-to-wall approach is used in the simulations without invoking any wall-function. In the reference DNS study\cite{ranjan2017new}, the value of $y_1^{+}$ is around 0.1 for most part of the blade.



\subsection{Turbulence Models}
\new{Table~\ref{tab:models} presents a list of all the turbulence models  employed for the current study, along with the underlying transport equations they solve as well as certain special features. Their performances for current transitional flow on T106A blade as described in subsequent sections are also summarized. Specifically the results are reported for the following turbulence models: Spalart-Allmaras (SA), Realizable $k$-$\epsilon$ and Shear-Stress Transport (SST) $k$-$\omega$; and the four-equation Langtry-Menter (LM) model that includes transition. Here $k$ and $\epsilon$ are turbulent kinetic energy and its dissipation rate respectively;  $\omega$ represents a turbulent frequency of the flow. These models illustrate superior performance in separated flows and therefore have been used for this study.  In the four equation Langtry-Menter model\cite{menter1994two},  transport equations are solved for $k$, $\omega$, intermittency $\gamma$ and transition momentum thickness Reynolds number $Re_{\theta}$.} 

Simulations were also performed with the three-equation $k$-$\epsilon$-$R_t$ model due to Goldberg \cite{goldberg2009k}, as well as the nonlinear explicit algebraic Reynolds stress model (EARSM) with a curvature correction developed by Hellsten \cite{hellsten2005new}. The results are similar to or less satisfactory than those of the turbulence models reported here and hence are not included in the discussion for brevity.

\begin{table} 
\begin{center}
      \begin{threeparttable}    
    \caption{Models employed in current simulations}
    \label{tab:models}
    \centering
\begin{tabular}{|c|c|c|c|}
\toprule
\new{\textbf{Model}}\tnote{1} & \textbf{Transport}  & \textbf{Special}  & \textbf{T106A} \\
 &  \textbf{equations} & \textbf{Features} & \textbf{Performance}\\
\midrule
    SA\cite{spalart1992one}    & $\Tilde{\nu}$ & Preferred for wall-bounded aerodynamic flows & Unable to predict SB\tnote{2}  \\
    \hline
    SST\cite{menter1994two} & $k$,$\omega$& Low-$Re$ turbulence model & -do- \\
        \hline
    RKE\cite{shih1994new}& $k$, $\epsilon$ & Realizability constraint in $\epsilon$ & -do- \\
        \hline
    EARSM\cite{hellsten2005new} & $k$,$\omega$ & Nonlinear $\tau_{ij}$ formulation & -do- \\
        \hline
    KERT\cite{goldberg2009k} & $k$,$\epsilon$,$R$ &  No inflow turbulence decay in
external flows& -do-\\
        \hline
    LM\cite{langtry2009correlation}  &   $k$,$\omega$,$\gamma$,$Re_{\theta}$  & $\gamma-Re_{\theta}$ equations to describe transition & Predicts a short SB\\
        \hline
    LNS\cite{batten2002lns} &$k$,$\epsilon$, $\bar{k}$ & Embedded hybrid RANS/LES   &Unable to predict SB\\
    \hline
    LNS-LEST\cite{batten2004interfacing} &$k$,$\epsilon$, $\bar{k}$ & LNS with Modeled$\leftrightarrow$Resolved fluctations    &Accurately predicts SB \\
    \bottomrule
\end{tabular}
     \begin{tablenotes}
     \item[1] SA: Spalart-Allamaras, SST: Shear-stress transport, RKE: Realizable $k$-$\epsilon$, EARSM: Explicit algebraic Reynolds stress model, KERT: $k$-$\epsilon$-$R_t$,LM: Langtry-Menter, LNS: Limited numerical scales, LNS-LEST: LNS with Large-Eddy Stimulation.\\
     \item[2] Separation bubble.
   \end{tablenotes}
    \end{threeparttable}%
    \end{center}
\end{table}

In the LNS model, the cubic $k$-$\epsilon$  model is used for the RANS calculations except in the regions where the mesh is sufficiently resolved and the model switches to a Smagorinsky-model based LES. The blending between RANS and LES is achieved by damping the modeled stress tensor as:
\begin{equation}\label{eq:lns}
\displaystyle    \tau_{LNS} = \alpha ~ \tau_{RANS}
\end{equation}
where $\tau_{RANS}$ is the solution obtained using the underlying RANS model, $\alpha$ is a latency parameter that is a function of 
turbulent length and velocity scales adopted for the eddy viscosity in both the RANS and the LES models,
\begin{equation}\label{eq:alpha}
\displaystyle    \alpha = \operatorname{min}\left(\frac{\nu_t^{LES},\nu_t^{RANS}}{\nu_t^{RANS}}\right)
\end{equation}
where $\nu_t^{RANS}$ and $\nu_t^{LES}$ are the effective eddy viscosities obtained using the cubic $k$-$\epsilon$ and subgrid scale (SGS) models respectively. Smagorinsky model is selected as SGS model for the LES component, such that-
\begin{equation}
\displaystyle    \nu_t^{LES} = C_s \Delta^2 S 
\end{equation}
where $C_s$ is the Smagorinsky coefficient, taken as 0.05, $S$ is the magnitude of the strain-rate tensor. $\Delta$ is the filter length used to distinguish between unresolvable (modeled) and resolvable scales of motion, given as:
\begin{equation}
\displaystyle    \Delta = 2 \operatorname{max} [\Delta x,~\Delta y,~\Delta z]
\end{equation}
Therefore, on sufficiently refined grids, $\nu_t^{LES} < \nu_t^{RANS}$ which gives $\alpha < 1$ from equation~\ref{eq:alpha}. 
Therefore,  from equation~\ref{eq:lns}, $\tau_{RANS}$ is reduced to LES-like values.  Detailed description of this implementation can be found in Batten \textit{et al.}\cite{batten2002lns}. 

An example of operating regions for RANS and LES in a T106 computational domain is shown in Labourasse and Sagaut\cite{labourasse2004advance} for a zonal LES/RANS approach. Specifically, the LES equations are solved in the aft region of the suction side where the transition occurs and the flow becomes unsteady. LNS is not a zonal approach, i.e. the physical boundaries between RANS and LES are not predefined, but as shown earlier, a locally refined mesh ensures that stresses are scaled-down to LES-like values in this region. Resolving scales with LES will be very crucial in the expected unsteady region (see Fig.~\ref{fig:domain_lns}(a)).   


\subsection{Energy Transfer between RANS/LES interfaces}
The steady or slowly evolving RANS solution  does not provide any turbulent fluctuations, therefore the transfer of data between RANS and LES remains an issue for most implementations of hybrid RANS/LES approaches\cite{spalart1997comments, labourasse2004advance}.
For proper unsteady boundary conditions, the statistical energy from RANS needs to be converted to directly-resolved kinetic energy in LES. 
In its absence, the potentially resolvable turbulence kinetic energy can be damped at blend regions because of the different effective
viscosity (see eqn.~\ref{eq:alpha}), and hence is likely to exhibit a fluctuation deficit.  This is not very critical in flows with  large inherent unsteadiness, which are self-sustaining  as they automatically generate resolved-scale disturbances which do not damp.
However, for transitional flows in low pressure turbine blades where there is only a thin separation, this remains a concern as shown in the subsequent sections in LNS simulations without any external stimulation i.e. TKE is allowed to decay at the local dissipation rate. 

In order to address this issue for the current transitional flow on the blade, we use the Large-Eddy STimulation (LEST) approach as proposed by Batten \textit{et al.}\cite{batten2004interfacing}. 
This approach is based on the concept of transfer of turbulence energy between scales in a statistical sense as represented in the energy spectrum. 
Turbulence kinetic energy in LNS includes both resolvable (large-scale) and modeled (small-scale) components of energy. 
On a locally refined mesh, the large-scale, statistically represented energy is converted to resolved-scale energy using a synthetic turbulence approach, so that they do not dissipate artificially and therefore do not bypass the energy cascade. 

Specifically,  an anisotropic fluctuating field is created using a scale similarity argument by superimposing white noise on the mean field, while ensuring realistic spatial and temporal correlations. As input, it requires the local stress tensor and the length and time scales of the turbulence, which are  directly available from the RANS field solved using the cubic $k$-$\epsilon$ model.

\section{Results \& Discussion}
In this section, the spanwise averaged flow characteristics
are presented.
The results from the current study are compared extensively against experimental data\cite{stadtmuller1} as well as recent DNS\cite{ranjan2017new} results. 
The DNS results are obtained in a compressible framework using the unstructured finite-volume code ANUROOP on a highly resolved grid (160 million grid points), and have an excellent match with the experiments.       



\begin{figure}
\centering
\subfloat[Pressure distribution. DNS data is from Ranjan \textit{et al.}\cite{ranjan2017new} on a 161 million grid.]{\includegraphics[width=.9\textwidth, trim={0.8cm 0cm 2cm 1cm},clip]{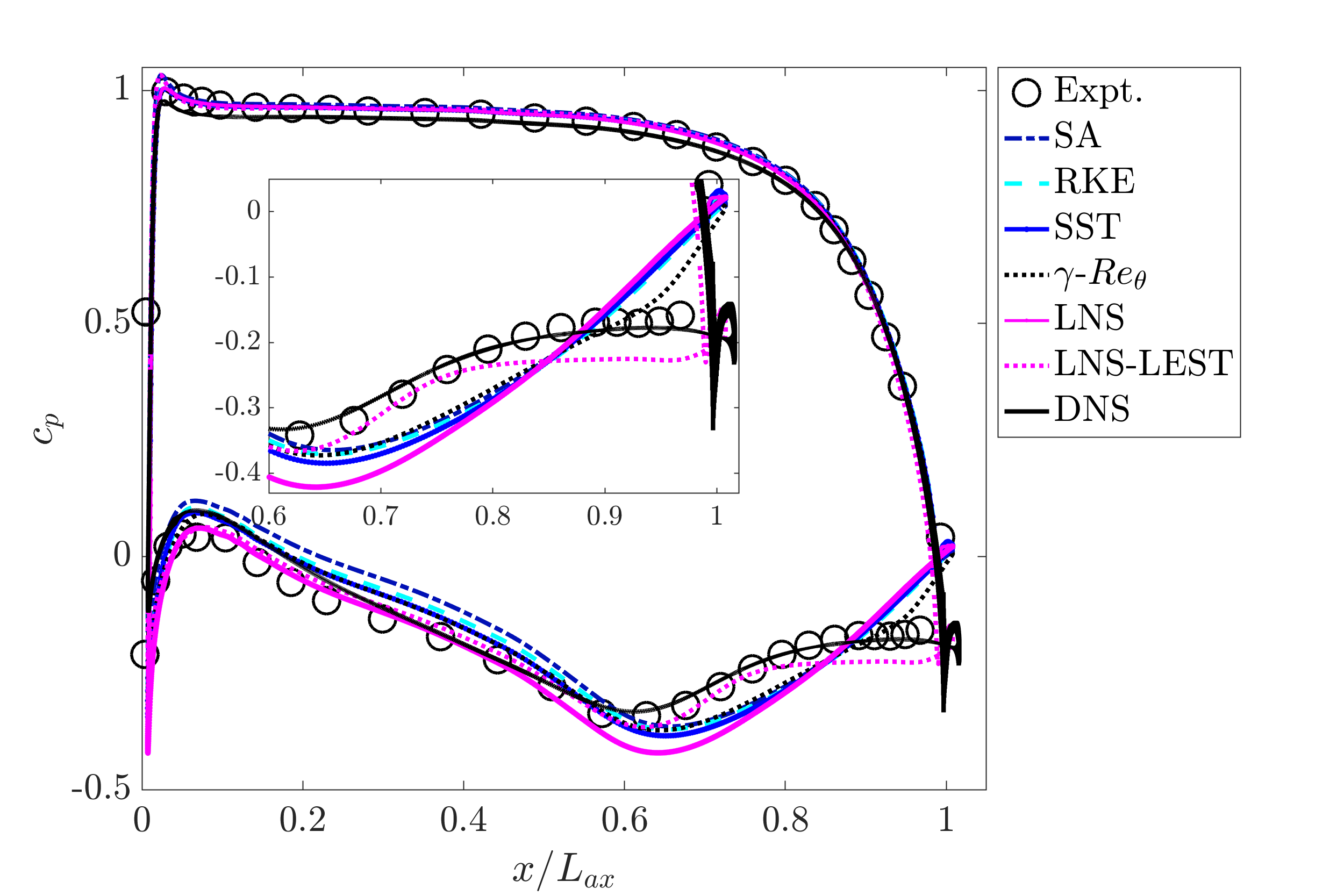}}\\
\subfloat[Isentropic Mach number distribution.]{\includegraphics[width=.9\textwidth, trim={0.8cm 0cm 2cm 1cm},clip]{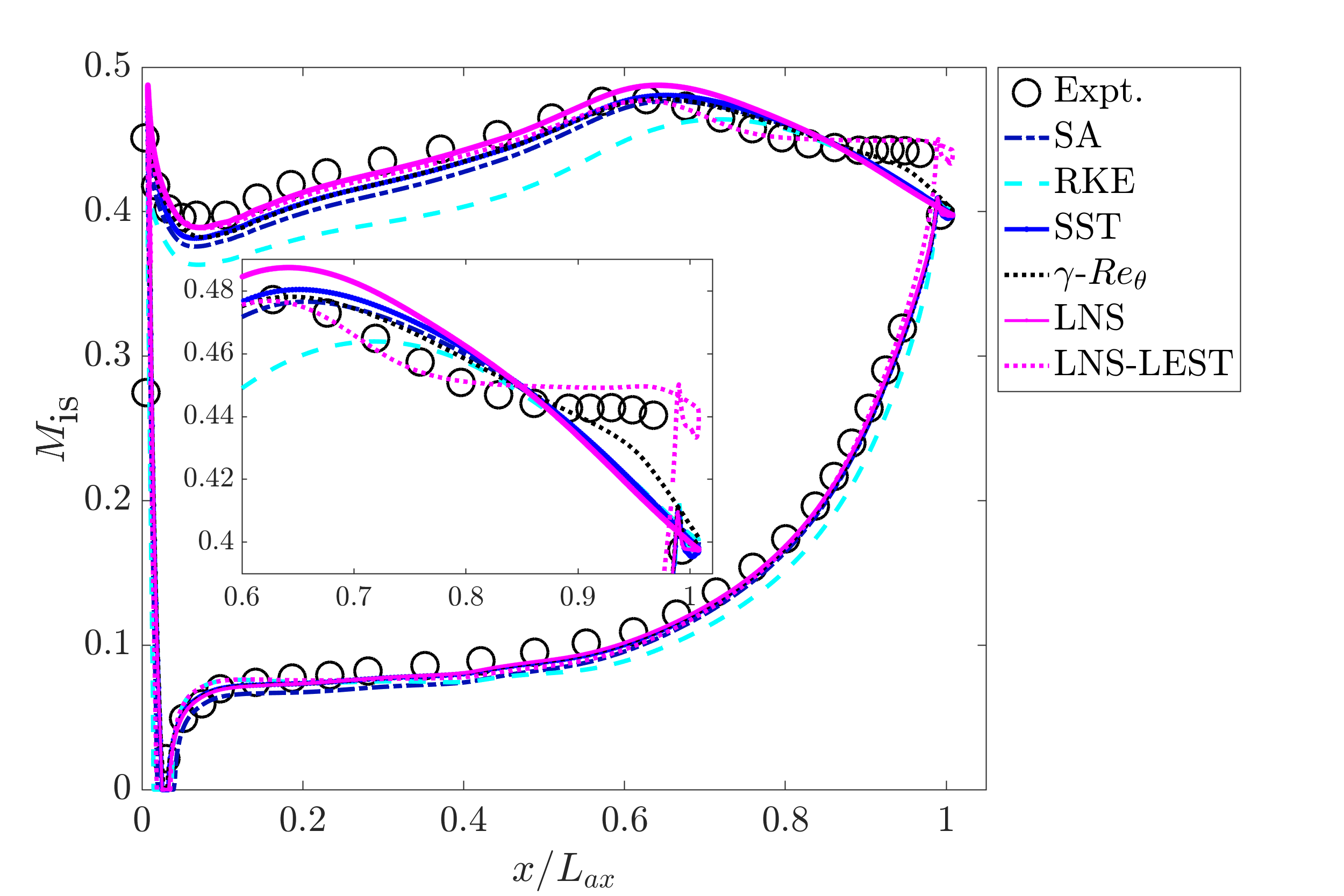}}
\caption{Mean flow on the T106A blade. In the inset, a zoomed view near the trailing edge of the suction side is shown. All the RANS models are unable to capture the pressure plateau in this region. Predictions by LNS with large-energy stimulation (LNS) are very close to the experiments.}
\label{fig:cp_rans}
\end{figure}

The flow over the T106A is first analyzed using the surface pressure distribution for which experimental data at $Re = 51,831$ is available from Stadtmuller\cite{stadtmuller1}. This quantity  is expressed by the pressure co-efficient 
\begin{align}
    \displaystyle    c_p = \frac{p-p_2}{p_{t1}-p_2} 
\end{align}
where $p_{t1}$ and $p_2$ are the total pressure at the inlet and back pressure at the outlet respectively. Further, the loading on the blade is also calculated through the isentropic Mach number defined as:
\begin{equation}
    \displaystyle M_{is} = \sqrt{\frac{2}{\gamma-1}\left[\left(\frac{p_{t1}}{p}\right)^{\frac{\gamma-1}{\gamma}}-1\right]}
\end{equation}
The $c_p$ and $M_{is}$ distributions along the axial chord are shown in Fig.~\ref{fig:cp_rans}(a) and (b) respectively for all the models tested as well as the results from the experiment\cite{stadtmuller1}. The $c_p$ results from high resolution DNS with 161 million grid is plotted for comparison.  The flow on the pressure side remains laminar and is predicted well by all the models, as seen in both $c_p$ and $M_{is}$ plots.
On the other hand, the flow field on the suction side is quite complex. 
Due to leading edge acceleration, there is a region of strong favorable pressure gradient till $x/L_{ax}=0.6$, where there is a peak. All the models capture this region well.
The boundary layer starts  decelerating beyond this point due to an adverse pressure gradient, and in this region the predictions of the models differ.

Both DNS and experiment show a plateau in this region which seems an indication of a dead air region due to separation.  
All the turbulence models miss this plateau except for the  Langtry-Menter  $\gamma$-$Re_{\theta}$ transition model where a small kink is observed in the pressure profile near $x/{L_{ax}}=0.94$. 
This kink is likely to be a manifestation of a `short' separation bubble \cite{marxen2007numerical}, as will be discussed later, and has only a local effect as the pressure distribution largely remains unaffected. 
Further, as the area inside the $C_p$ in Fig.~\ref{fig:cp_rans}(a) represents the relative
amount of the blade loading, all the turbulence models fail to predict the losses due to the separation found in DNS and experimental studies.
This observation is not unexpected as there are several studies in the literature that have reported the inability of turbulence models in predicting separation and transition.
Particularly for flow on the T106 blade at sub-critical conditions, none of the studies so far has achieved significant success, to the best of our knowledge, with just the turbulence models.
In the recent study by Akolekar \textit{et al.}\cite{akolekar2019development}, transition models LKE and $k\bar{v_2}\omega$ do predict the separation bubble, albeit slightly longer and thicker than their DNS. 
Their study with the Langtry-Menter transition model however does not show any separation.

We now focus on the results from hybrid RANS/LES simulations. The findings from LNS - with and without large-energy stimulation, marked by LNS and LNS-LEST respectively, are interesting. While LNS without stimulation does not show any plateau and largely follows the curve as predicted by other turbulence models, the results with LNS-LEST are very close to the experiments. 

\begin{figure}
\centering
\subfloat[LNS]{\includegraphics[trim=0.1cm 0.1cm 0.1cm 0.1cm, clip, width=0.74\linewidth]{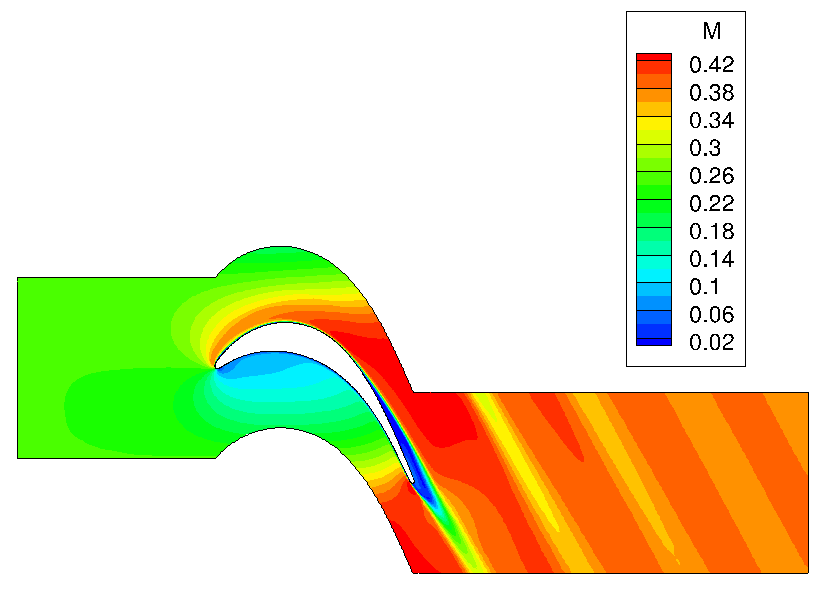}}\\
\subfloat[LNS-LEST]{\includegraphics[trim=0.1cm 0.1cm 0.1cm 0.1cm, clip, width=0.74\linewidth]{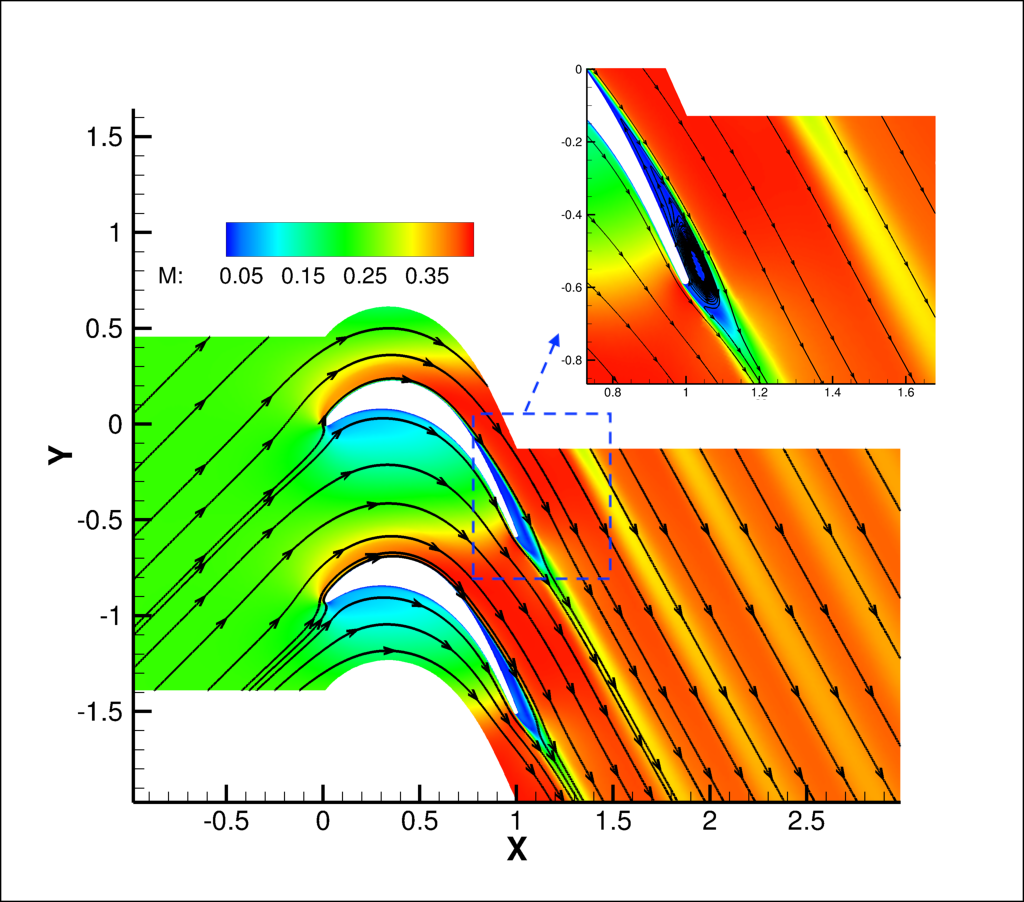}}
  \caption{Flow predicted using LNS model without (a) and with (b) stimulation of large eddy using statistical energy. All lengths are normalized by $L_{ax}$.}
  \label{fig:lns2}
\end{figure}


In order to further assess the effect of energy stimulation on the cascade flow, we show the Mach number contours in Fig.~\ref{fig:lns2} for simulations performed with and without stimulation. The streamlines are also plotted. The contours on the pressure side are very similar to each other. However, on the suction side, a slightly higher velocity magnitude in the outer portion of the boundary layer is seen for the LEST case despite the same distribution on the wall (Fig.~\ref{fig:cp_rans}). Near the trailing edge, the difference becomes very significant as a thin but long dead air region shown by very low Mach number, is observed for the LNS-LEST case, while the boundary layer remains attached for LNS without stimulation.

It should be reemphasized that LEST provides unsteady boundary conditions for the embedded LES region by spatially and temporally correlated synthetic reconstruction of the statistical turbulence energy obtained from the solution in the RANS region.
In the absence of the LEST model or any sufficiently inherent large unsteadiness, the equations will change from RANS to LES in resolved grids but not the solution data, which may result in a nonphysical drop in the time-averaged total stresses due to dissipation. 

As in the present case the boundary layer is laminar for a large part of the blade, there is not enough unsteadiness in the embedded LES regions to sustain turbulence. In the absence of any external stimulation, the statistical energy obtained from RANS gets quickly laminarized in LES. Therefore, the LNS without stimulation shows a completely laminar boundary layer. This behavior is observed in several hybrid simulations~\cite{baggett1998feasibility,nikitin2000approach}. However, in the LEST case, the  smaller scale perturbations created using statistical energy successfully maintain large-scale unsteady motion in the outer portion of the boundary layer, thereby preventing the laminarization in the LES core region due to sudden suppression of unresolved stresses, and thus predicting the right flow physics. LEST has been found to be successful in predicting such behavior for many complex flows otherwise very challenging\cite{mainenti2010comparative, batten2004interfacing}, but its effectiveness in predicting the  transitional flow on the LPT blade has not been shown before. Another feature of the LEST prediction on the LPT blade in the cascade can be seen in Fig.~\ref{fig:lns2}(b) in the form of periodic wake structures.

\begin{figure}
\centering
\includegraphics[trim=0 0 0 0, clip, width=0.9\linewidth]{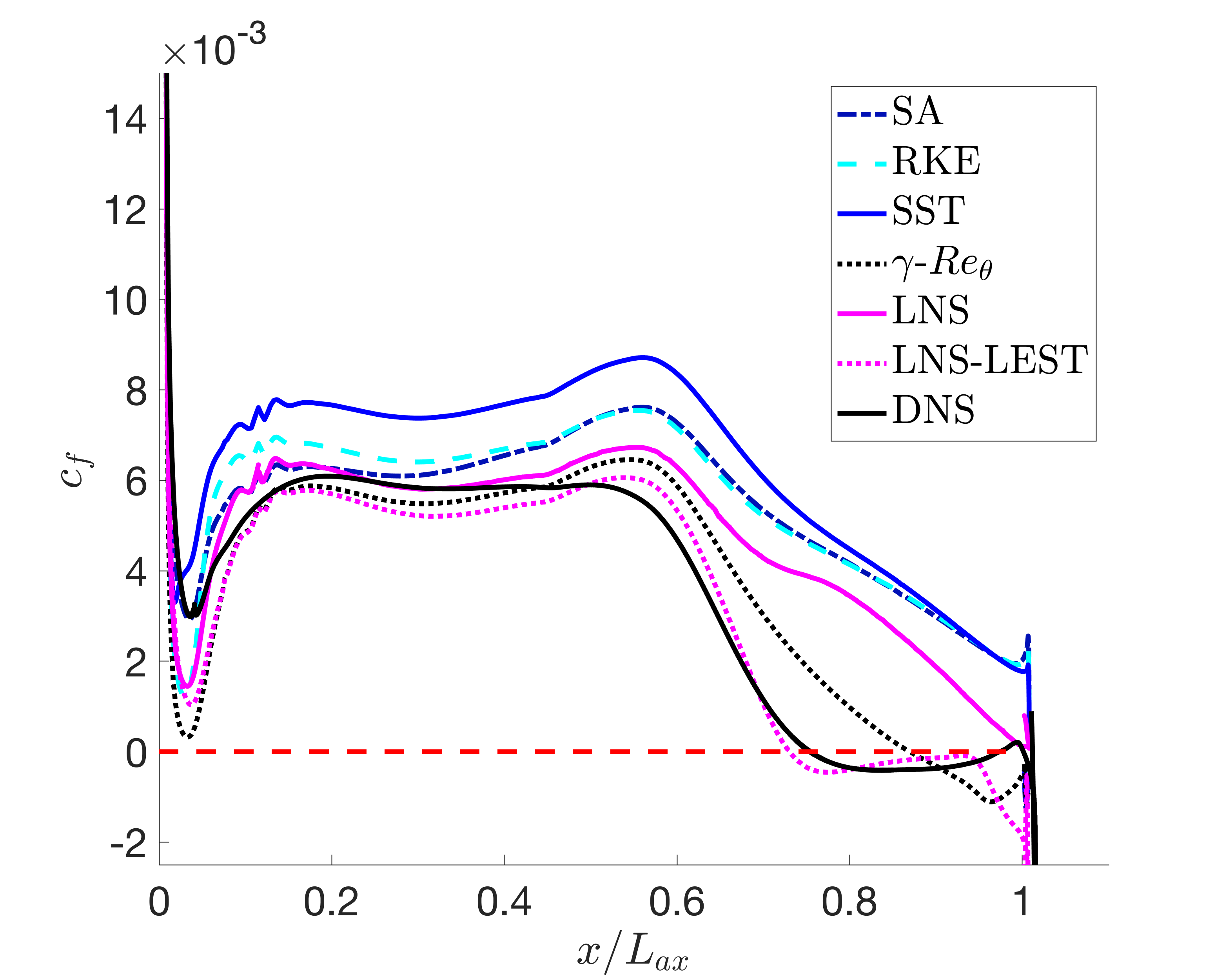}
\caption{Skin-friction distribution on the blade. DNS data is from Ranjan \textit{et al.}\cite{ranjan2017new} on a 161 million grid.}
\label{fig:cf_rans}
\end{figure}

We return to the description of flow behavior on the suction side where a plateau close to the trailing edge was observed. Stadtmuller\cite{stadtmuller1} has attributed this plateau to the presence of a separation bubble, however, the extent and characteristics of this bubble could not be confirmed in that experiment with surface pressure and hot film measurements. 
Recent high-resolution DNS\cite{ranjan2015compressible,ranjan2017new} study describes the characteristics of this bubble in detail.   
Therefore, we compare the results from current simulations with those obtained from the DNS. First we compare the skin-friction coefficient $c_f$, defined as
\begin{equation}
\displaystyle  c_f = \frac{\tau_w}{\frac{1}{2}\rho_{\infty} u_{\infty}^2}
\end{equation}
where $\displaystyle \tau_w = \mu \frac{\partial u}{\partial y}|_{y=0}$ is the wall shear stress, quantitatively identifies the region of flow separation. This parameter is positive for the attached flow, and the point where it first becomes zero can be identified as the onset of separation. This quantity is plotted in Fig~\ref{fig:cf_rans} on the suction side for all the models considered as well as the DNS results. 

The leading edge on the suction side does not show any separation for any simulation. However, in the adverse pressure gradient region, a flow separation is observed in the DNS study that corresponds to the pressure plateau in the experiments (Fig.~\ref{fig:cp_rans}). As expected, the turbulence models do not predict this behavior except for the Langtry-Menter model, which predicts the onset of separation but much further downstream of the DNS results.  The prediction from LNS-LEST is however very close to that from DNS, in terms of both the onset of separation and separation length. The behavior of LNS without stimulation is very close to the RANS results without transition.  

\begin{figure}
\centering
\subfloat[SST]{\includegraphics[trim=0 0 0 0, clip, width=0.45\linewidth]{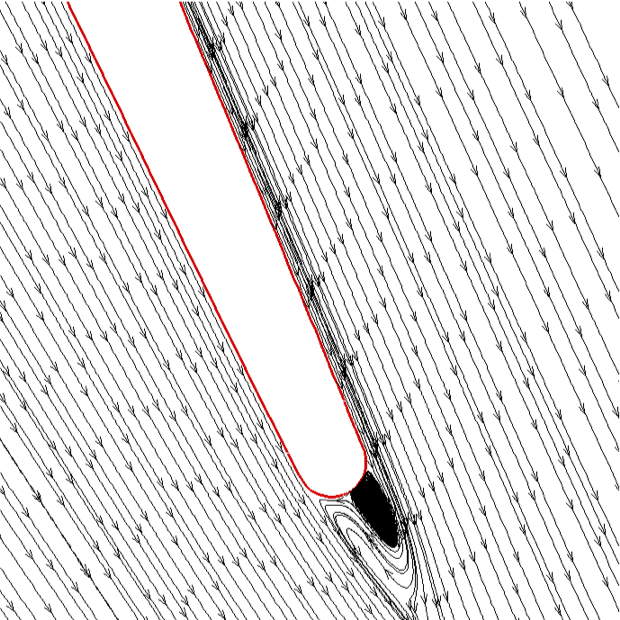}}\enskip
\subfloat[$\gamma$-$Re_{\theta}$]{\includegraphics[trim=2cm 0.5cm 0 0, clip, width=0.45\linewidth]{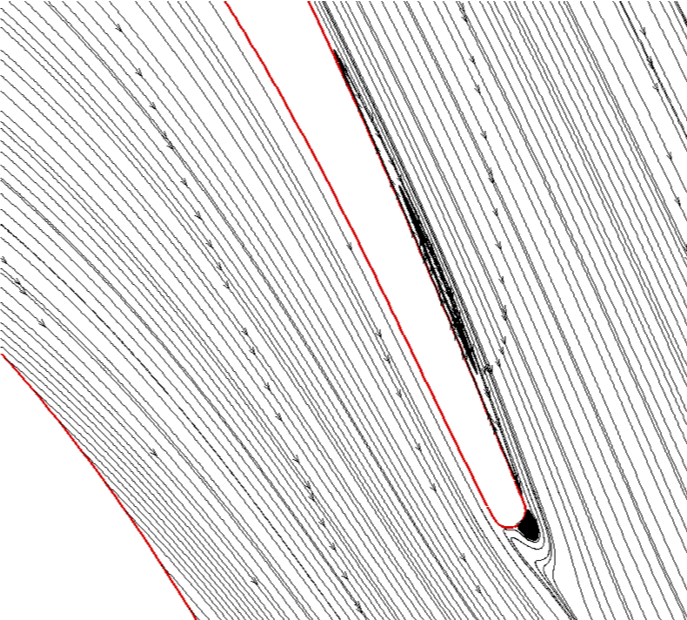}}\\
\subfloat[LNS-LEST]{\includegraphics[trim=0 1cm 1cm 1cm, clip, width=0.45\linewidth]{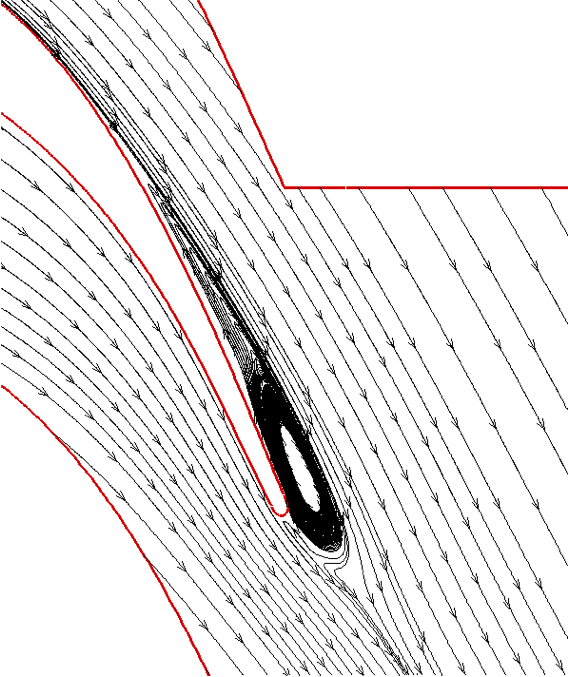}}\enskip
\subfloat[DNS]{\includegraphics[trim=1cm 0 3cm 0cm, clip, width=0.45\linewidth]{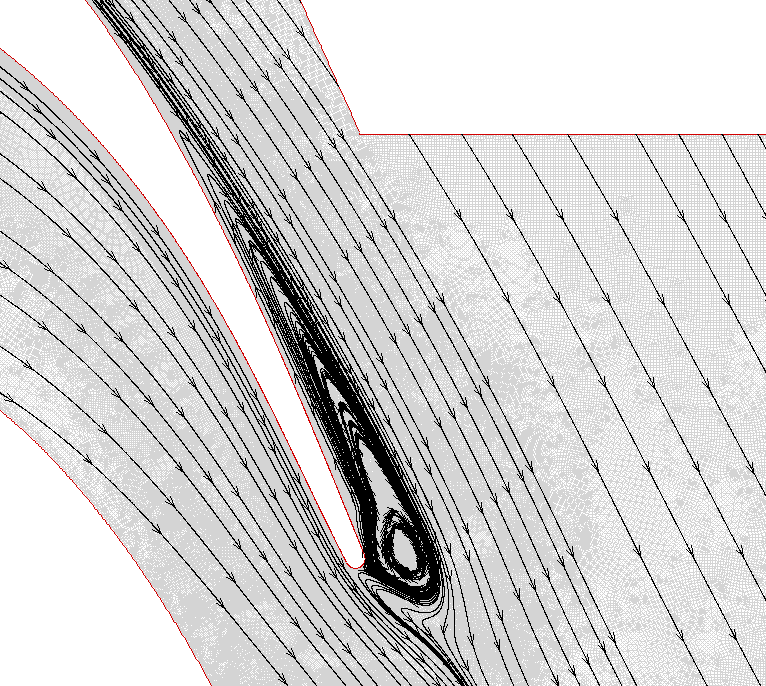}}
\caption{Trailing-edge Separation Bubbles}
\label{fig:separation_rans}
\end{figure}

In order to characterize the separation region, we plot the surface streamlines on the midplane of the domain. Figure~\ref{fig:separation_rans} shows these plots for SST and Langtry-Menter models, LNS-LEST and DNS. The streamlines of other turbulence models as well as LNS are very similar to the predictions of SST, and hence are not shown here for brevity. 

In the case of SST in Fig.~\ref{fig:separation_rans}(a), a small recirculation region is seen in the wake region of the blade in an otherwise attached flow. In the Langtry-Menter simulation, a very thin separation bubble is observed near the trailing edge but this bubble does not affect the shape of streamlines significantly or alter the blade exit angle. Limited and localized effects of the bubble along with a very small plateau in wall pressure due to it are the features of a `short' separation bubble as categorized in reference\cite{marxen2011effect}. On the other hand, the structure of separation bubbles and the location of the separation point in the DNS and LNS-LEST results are very similar, and are more pronounced than the short separation bubble in the Langtry-Menter model. Further, consistent with the skin-friction observations shown in Fig.~\ref{fig:cf_rans}, the separation occurs earlier in the former cases than in the latter. The onset of separation $x_s$, length of bubble $L_s$ and the maximum thickness $h$ of these separation bubbles are given in Table~\ref{tab:sep}. The value of $h$ in DNS and LNS-LEST are very close, and an order of magnitude higher than that in the short separation bubble in Fig.~\ref{fig:separation_rans}(b). Further, these separation bubbles greatly displace the streamlines and have higher exit flow angles from the design value, leading to higher losses. Therefore, these bubbles can be classified under `long' separation bubbles.      

\begin{table}[]
    \centering
\begin{tabular}{|c|c|c|c|}
\toprule
\textbf{Parameter} & $x_s$ & $L_s$ & $h$ \\
\midrule
DNS & 0.755 & 0.345 & 0.07  \\
\hline
$\gamma$-$Re_{\theta}$ & 0.87 & 0.13 & 0.007  \\
\hline
LNS-LEST & 0.733 & 0.367  & 0.068  \\
    \bottomrule
\end{tabular}
    \caption{Characterization of Separation Bubbles}
    \label{tab:sep}
\end{table}

\begin{figure}
\centering
\subfloat[$\gamma$-$Re_{\theta}$]{\includegraphics[trim=1cm 1cm 3cm 2cm, clip, width=0.48\linewidth]{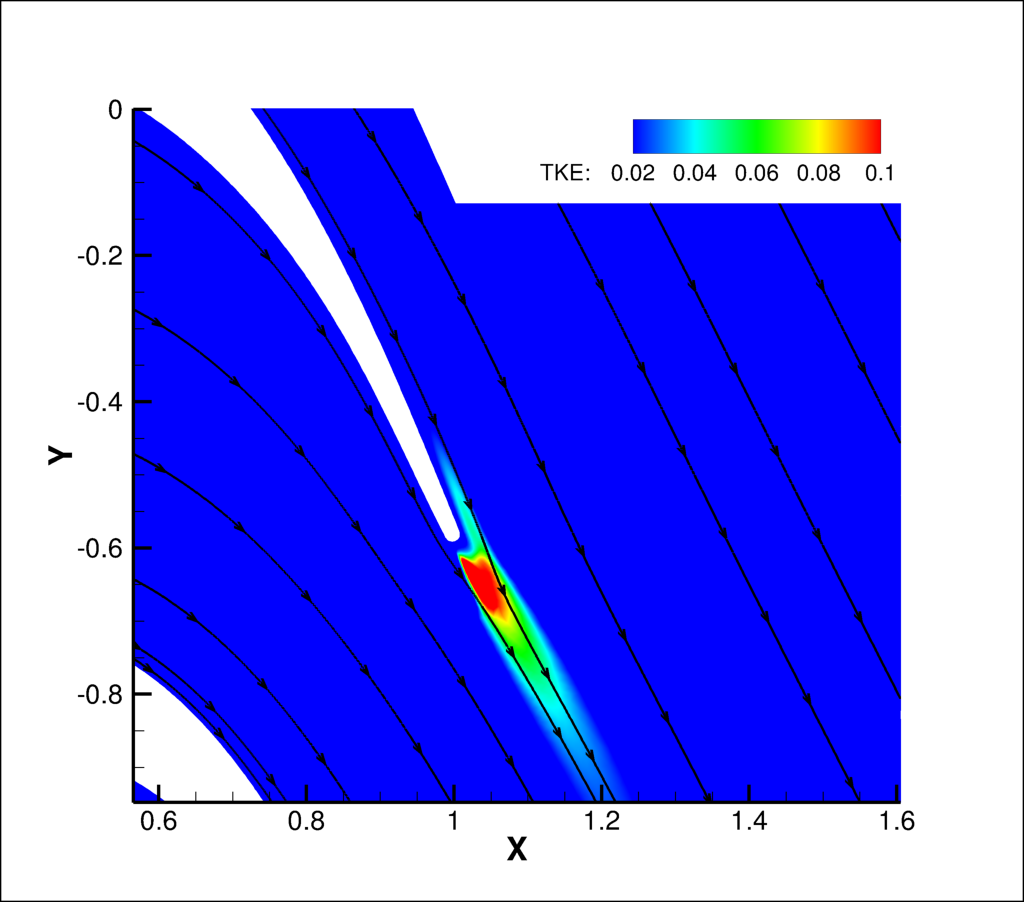}}\enskip
\subfloat[LNS-LEST]{\includegraphics[trim=1cm 1cm 3cm 2cm, clip, width=0.48\linewidth]{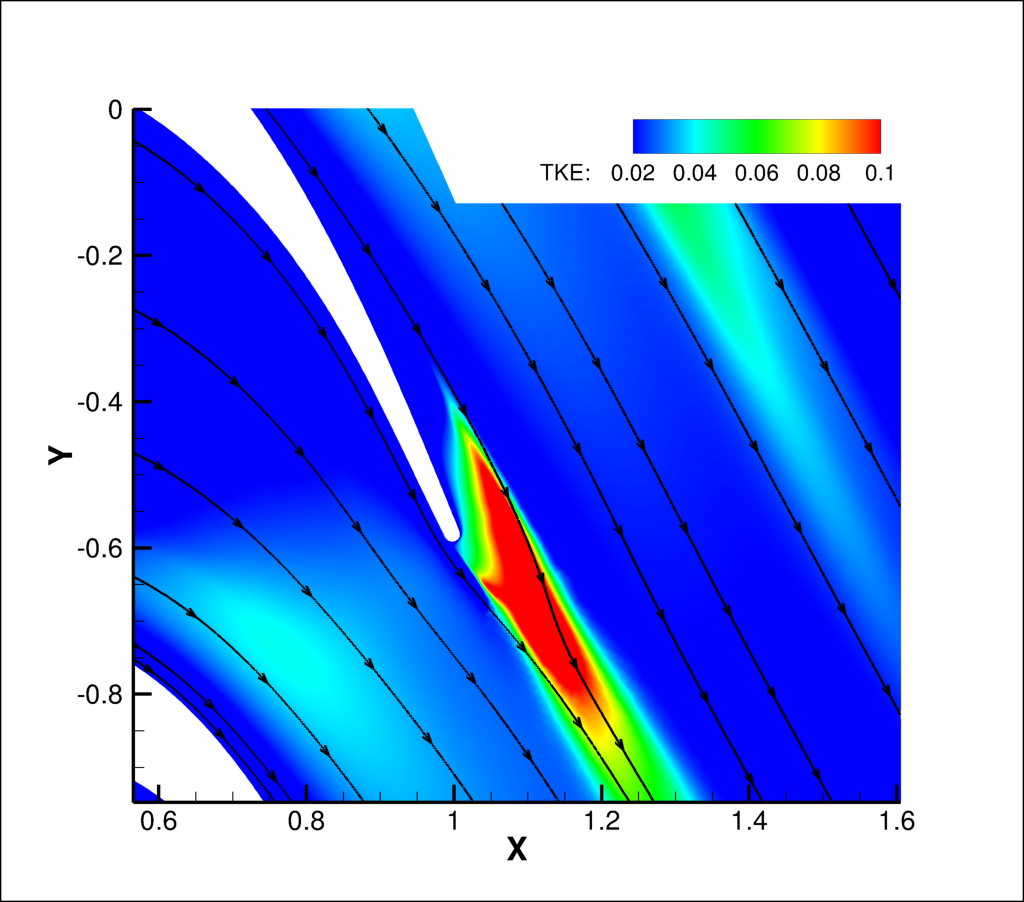}}
\caption{Turbulent kinetic energy (TKE) in the wake of the blade. All lengths are normalized by $L_{ax}$.}
\label{fig:tke_rans}
\end{figure}

Lastly, we show the distribution of mean turbulent kinetic energy (TKE) $k$ downstream of the blade for both $\gamma$-$Re_{\theta}$ and LNS-LEST simulations (Fig.~\ref{fig:tke_rans}). 
The increase in TKE downstream of the onset of separation shows that a laminar separation bubble triggers transition. 
Briefly, the separated shear layer is very unstable and triggering of Kelvin-Helmholtz instability leads to generation of turbulence.
This transition occurs roughly around $x/L_{ax}=0.95$ in both cases.
As found in Michelassi \textit{et al.}\cite{michelassi2002}, the length between the separation  and trailing edge is insufficient to have a fully turbulent boundary layer, and the peak in TKE is obtained  in the wake region. 
Further, the transition is more pronounced in LNS-LEST, and lasts longer in the wake compared to the Langtry-Menter case where it has only the local effects of a short separation bubble. 
It will be reasonable to conclude that the production of TKE in the Langtry-Menter model is much lower than that in the LNS-LEST case.
As remarked by Marciniak \textit{et al.}\cite{marciniak2010predicting} the transition model requires tuning to correct for the prediction of TKE, but it may need inputs from DNS and LES results.

\section{Concluding Remarks}
Transitional flow past the low-pressure turbine (LPT) blade T106A in a cascade is simulated using RANS and hybrid RANS/LES approaches. 
Because of the high angle of incidence and low $Re$, the flow condition is similar to that experienced by the turbine blade cruising at high altitude; i.e. an attached boundary layer at the leading edge and a separation bubble on the suction side due to adverse pressure gradient towards the trailing edge.    
Predicting such flows using turbulent and transitional models is challenging due to inherent difficulties in predicting weak unsteadiness \new{such as that found in thin separation regions}.

Among the six popular low-$Re$ turbulence models used here, none of them shows the ability to predict separation and hence transition. 
The Langtry-Menter $\gamma$-$Re_{\theta}$ transition model is moderately successful in predicting separation, however the onset of separation as well as the structure of the separation bubble are very different from what is obtained through DNS.
Further, the pressure distribution on the suction side differs significantly in this region from experimental results.

Hybrid RANS/LES simulations are performed with the embedded approach of using Limited Numerical Scales (LNS), with the same grid as in RANS.
In the absence of any model to convert a mean statistical energy into turbulent fluctuations, the turbulent stress gets damped in the LES  zone, and an attached boundary layer is obtained similar to those observed with turbulence models.
However, when synthetic turbulence with correct statistical characteristics, that gives resolved-scale energy in the locally fine mesh, is provided to the embedded LES zone, LNS is able to predict separation. 
The solution thus recovered are very close to DNS and experimental results.
This strongly suggests that in hybrid simulations for transitional flow on LPT blades, a model which provides an automatic unsteady flow for the LES zone is necessary to predict separation in an otherwise attached boundary layer. 

The separation bubble predicted by the $\gamma$-$Re_{\theta}$ model may be classified as a relatively `short' bubble compared to the longer bubbles obtained in DNS and LNS studies. This is because the transition model does not predict the onset of separation accurately, which further may be due to underestimation of production of turbulent kinetic energy in this model. The prediction can be improved by further calibrating the parameters for attached boundary layers with thin separation, but this may need information from a truth model such as an experiment or a well resolved LES or DNS.

Finally, we conclude that hybrid RANS/LES, with appropriate mechanism of representing turbulence kinetic energy on resolved and unresolved scales,  shows promise as a reliable tool for simulating boundary layers on turbine blades with reasonable RANS-like grids.
The effect of upstream wakes on turbine blades presents additional challenges for these models due to interaction between boundary layers and wakes.
Our current efforts are addressing this issue.

\bibliographystyle{asmems4}

\bibliography{refer_jot}

\begin{thebibliography}{10}

\bibitem{stadtmuller1}
Stadtmuller, P., 2002.
\newblock Investigation of wake-induced transition on the {LP} turbine cascade
  {T106A-EIZ}.
\newblock Tech. rep.

\bibitem{hodson2007role}
Hodson, H., and Howell, R., 2007.
\newblock ``The role of transition in high lift low pressure turbines''.
\newblock {\em Lecture Series-Von Karman Institute for Fluid Dynamics, {\bf
  2}}, p.~3.

\bibitem{schobeiri2014comparative}
Schobeiri, M.~T., and Nikparto, A., 2014.
\newblock ``A comparative numerical study of aerodynamics and heat transfer on
  transitional flow around a highly loaded turbine blade with flow separation
  using {RANS}, {URANS} and {LES}''.
\newblock In ASME Turbo Expo 2014: Turbine Technical Conference and Exposition,
  American Society of Mechanical Engineers Digital Collection.

\bibitem{kalitzin2003dns}
Kalitzin, G., Wu, X., and Durbin, P.~A., 2003.
\newblock ``{DNS} of fully turbulent flow in a {LPT} passage''.
\newblock {\em International Journal of Heat and Fluid Flow, {\bf 24}}(4),
  pp.~636--644.

\bibitem{sandberg2015compressible}
Sandberg, R.~D., Michelassi, V., Pichler, R., Chen, L., and Johnstone, R.,
  2015.
\newblock ``Compressible direct numerical simulation of low-pressure
  turbines---part {I}: Methodology''.
\newblock {\em Journal of Turbomachinery, {\bf 137}}(5), p.~051011.

\bibitem{ranjan2017new}
Ranjan, R., Deshpande, S., and Narasimha, R., 2017.
\newblock ``New insights from high-resolution compressible {DNS} studies on an
  {LPT} blade boundary layer''.
\newblock {\em Computers \& Fluids, {\bf 153}}, pp.~49--60.

\bibitem{michelassi2002}
Michelassi, V., Wissink, J., and Rodi, W., 2002.
\newblock ``Analysis of {DNS} and {LES} of flow in a low pressure turbine
  cascade with incoming wakes and comparison with experiments''.
\newblock {\em Flow, Turbulence and Combustion, {\bf 69}}, pp.~295--329.

\bibitem{wauters2018study}
Wauters, J., and Degroote, J., 2018.
\newblock ``On the study of transitional low-{R}eynolds number flows over
  airfoils operating at high angles of attack and their prediction using
  transitional turbulence models''.
\newblock {\em Progress in Aerospace Sciences, {\bf 103}}, pp.~52--68.

\bibitem{dick2017transition}
Dick, E., and Kubacki, S., 2017.
\newblock ``Transition models for turbomachinery boundary layer flows: a
  review''.
\newblock {\em International Journal of Turbomachinery, Propulsion and Power,
  {\bf 2}}(2), p.~4.

\bibitem{batten2002lns}
Batten, P., Goldberg, U., and Chakravarthy, S., 2002.
\newblock ``{LNS} - an approach towards embedded {LES}''.
\newblock In 40th AIAA Aerospace Sciences Meeting \& Exhibit, p.~427.

\bibitem{frohlich2008hybrid}
Fr{\"o}hlich, J., and Von~Terzi, D., 2008.
\newblock ``Hybrid {LES}/{RANS} methods for the simulation of turbulent
  flows''.
\newblock {\em Progress in Aerospace Sciences, {\bf 44}}(5), pp.~349--377.

\bibitem{labourasse2004advance}
Labourasse, E., and Sagaut, P., 2004.
\newblock ``Advance in {RANS}-{LES} coupling, a review and an insight on the
  {NLDE} approach''.
\newblock {\em Archives of Computational methods in Engineering, {\bf 11}}(3),
  pp.~199--256.

\bibitem{spalart1997comments}
Spalart, P.~R., 1997.
\newblock ``Comments on the feasibility of {LES} for wings, and on a hybrid
  {RANS}/{LES} approach''.
\newblock In Proceedings of {F}irst AFOSR {I}nternational {C}onference on
  {DNS}/{LES}, Greyden Press.

\bibitem{Chakravarthy2002}
Chakravarthy, S.~R., Goldberg, U.~C., and Batten, P., 2002.
\newblock {\em Examples of Contemporary CFD Simulations}.
\newblock Springer Netherlands, Dordrecht, pp.~339--369.

\bibitem{jouvray2005computation}
Jouvray, A., and Tucker, P.~G., 2005.
\newblock ``Computation of the flow in a ventilated room using non-linear
  {RANS}, {LES} and hybrid {RANS}/{LES}''.
\newblock {\em International Journal for Numerical Methods in Fluids, {\bf
  48}}(1), pp.~99--106.

\bibitem{ranjan2015advances}
Ranjan, R., Deshpande, S., and Narasimha, R., 2016.
\newblock ``A high-resolution compressible {DNS} study of flow past a
  low-pressure gas turbine blade''.
\newblock In {\em Advances in Computation, Modeling and Control of Transitional
  and Turbulent Flows}. World Scientific, ch.~28, pp.~291--301.

\bibitem{marciniak2010predicting}
Marciniak, V., K{\"u}geler, E., and Franke, M., 2010.
\newblock ``Predicting transition on low-pressure turbine profiles''.
\newblock In V European Conference on Computational Fluid Dynamics (ECCOMAS CFD
  2010), Lisbon, Portugal, June, pp.~14--17.

\bibitem{akolekar2019development}
Akolekar, H., Weatheritt, J., Hutchins, N., Sandberg, R., Laskowski, G., and
  Michelassi, V., 2019.
\newblock ``Development and use of machine-learnt algebraic {R}eynolds stress
  models for enhanced prediction of wake mixing in low-pressure turbines''.
\newblock {\em Journal of Turbomachinery, {\bf 141}}(4).

\bibitem{Mayle1991}
Mayle, R.~E., 1991.
\newblock ``{The 1991 IGTI Scholar Lecture: The Role of Laminar-Turbulent
  Transition in Gas Turbine Engines}''.
\newblock {\em Journal of Turbomachinery, {\bf 113}}(4), 10, pp.~509--536.

\bibitem{lopez2016prediction}
Lopez, M., and Walters, D.~K., 2016.
\newblock ``Prediction of transitional and fully turbulent flow using an
  alternative to the laminar kinetic energy approach''.
\newblock {\em Journal of Turbulence, {\bf 17}}(3), pp.~253--273.

\bibitem{walters2008three}
Walters, D.~K., and Cokljat, D., 2008.
\newblock ``A three-equation eddy-viscosity model for {R}eynolds-averaged
  {N}avier--{S}tokes simulations of transitional flow''.
\newblock {\em Journal of Fluids Engineering, {\bf 130}}(12).

\bibitem{minot2016improvement}
Minot, A., Salah El-Din, I., Barrier, R., Boniface, J.-C., and Marty, J., 2016.
\newblock ``Improvement of laminar-turbulent transition modeling within a
  low-pressure turbine''.
\newblock In ASME Turbo Expo 2016: Turbomachinery Technical Conference and
  Exposition, American Society of Mechanical Engineers Digital Collection.

\bibitem{marciniak2015modeling}
Marciniak, V., 2015.
\newblock ``Modeling flows in low-pressure turbine cascades at very low
  {R}eynolds numbers''.
\newblock {\em CEAS Aeronautical Journal, {\bf 6}}(2), pp.~257--270.

\bibitem{lodefier2006modelling}
Lodefier, K., and Dick, E., 2006.
\newblock ``Modelling of unsteady transition in low-pressure turbine blade
  flows with two dynamic intermittency equations''.
\newblock {\em Flow, Turbulence and Combustion, {\bf 76}}(2), pp.~103--132.

\bibitem{muller2016dns}
M{\"u}ller, C., Baier, R.-D., Seume, J.~R., and Herbst, F., 2016.
\newblock ``{DNS}-based analysis of {RANS} predictions of a low-pressure
  turbine cascade''.
\newblock In ASME Turbo Expo 2016: Turbomachinery Technical Conference and
  Exposition, American Society of Mechanical Engineers Digital Collection.

\bibitem{nikparto2016numerical}
Nikparto, A., and Schobeiri, M.~T., 2016.
\newblock ``Numerical and experimental investigation of aerodynamics on flow
  around a highly loaded low-pressure turbine blade with flow separation under
  steady and periodic unsteady inlet flow condition''.
\newblock In ASME Turbo Expo 2016: Turbomachinery Technical Conference and
  Exposition, American Society of Mechanical Engineers Digital Collection.

\bibitem{chakravarthy1999cfd++}
Chakravarthy, S., 1999.
\newblock ``A unified-grid finite volume formulation for computational fluid
  dynamics''.
\newblock {\em International {J}ournal for {N}umerical {M}ethods in {F}luids,
  {\bf 31}}(1), pp.~309--323.

\bibitem{menter1994two}
Menter, F.~R., 1994.
\newblock ``Two-equation eddy-viscosity turbulence models for engineering
  applications''.
\newblock {\em AIAA Journal, {\bf 32}}(8), pp.~1598--1605.

\bibitem{goldberg2009k}
Goldberg, U., Peroomian, O., Batten, P., and Chakravarthy, S., 2009.
\newblock ``The $k$-$\varepsilon$-$r_t$ turbulence closure''.
\newblock {\em Engineering Applications of Computational Fluid Mechanics, {\bf
  3}}(2), pp.~175--183.

\bibitem{hellsten2005new}
Hellsten, A.~K., 2005.
\newblock ``New advanced kw turbulence model for high-lift aerodynamics''.
\newblock {\em AIAA journal, {\bf 43}}(9), pp.~1857--1869.

\bibitem{spalart1992one}
Spalart, P.~R., and Allmaras, S.~R., 1992.
\newblock ``A one-equation turbulence model for aerodynamic flows''.

\bibitem{shih1994new}
Shih, T.-H., Liou, W.~W., Shabbir, A., Yang, Z., and Zhu, J., 1994.
\newblock ``A new k-epsilon eddy viscosity model for high reynolds number
  turbulent flows: Model development and validation''.

\bibitem{langtry2009correlation}
Langtry, R.~B., and Menter, F.~R., 2009.
\newblock ``Correlation-based transition modeling for unstructured parallelized
  computational fluid dynamics codes''.
\newblock {\em AIAA {J}ournal, {\bf 47}}(12), pp.~2894--2906.

\bibitem{batten2004interfacing}
Batten, P., Goldberg, U., and Chakravarthy, S., 2004.
\newblock ``Interfacing statistical turbulence closures with large-eddy
  simulation''.
\newblock {\em AIAA Journal, {\bf 42}}(3), pp.~485--492.

\bibitem{marxen2007numerical}
Marxen, O., and Henningson, D., 2007.
\newblock ``Numerical simulation of the bursting of a laminar separation
  bubble''.
\newblock In {\em High Performance Computing in Science and Engineering’06}.
  Springer, pp.~253--267.

\bibitem{baggett1998feasibility}
Baggett, J.~S., 1998.
\newblock ``On the feasibility of merging {LES} with {RANS} for the near-wall
  region of attached turbulent flows''.
\newblock {\em Annual Research Briefs}, pp.~267--277.

\bibitem{nikitin2000approach}
Nikitin, N., Nicoud, F., Wasistho, B., Squires, K., and Spalart, P.~R., 2000.
\newblock ``An approach to wall modeling in large-eddy simulations''.
\newblock {\em Physics of Fluids, {\bf 12}}(7), pp.~1629--1632.

\bibitem{mainenti2010comparative}
Mainenti Leal~Lopes, V., dos Santos~Bonatto, A., Romano~Meneghini, J., and
  Saltara, F., 2010.
\newblock ``Comparative analysis of turbulence models for slat noise sources
  calculations employing structured meshes''.
\newblock In 16th AIAA/CEAS Aeroacoustics Conference, p.~3834.

\bibitem{ranjan2015compressible}
Ranjan, R., 2015.
\newblock ``Compressible dns studies of the boundary layer on a low pressure
  turbine blade at high incidence''.
\newblock PhD thesis, Jawaharlal Nehru Centre for Advanced Scientific Research.

\bibitem{marxen2011effect}
Marxen, O., and Henningson, D.~S., 2011.
\newblock ``The effect of small-amplitude convective disturbances on the size
  and bursting of a laminar separation bubble''.
\newblock {\em Journal of Fluid Mechanics, {\bf 671}}, pp.~1--33.

\end{thebibliography}













\section*{Acknowledgement}
We would like to thank the Gas Turbine Research Establishment (GTRE), Bangalore, for funding this project. We would also like to thank Dr. Sukumar Chakravarthy, METACOMP Technologies, USA for his generous support in providing CFD++ license and suggestions regarding this work.

%
%
\end{document}